# LiveDocs: Crafting Interactive Development Environments From Research Findings

Pedro Costa Klein[1],[*] Christoph Lehrenfeld[1], Markus Osterhoff[2] and Martin Uecker[3]

[1] Institute for Numerical and Applied Mathematics, University of Göttingen, Germany
[2] Institute for X-ray Physics, University of Göttingen, Germany
[3] Institute of Biomedical Imaging, TU Graz, Austria

**Abstract**

*Open Science is a recurrent topic in scientific discussion, and there is a current effort to make research more accessible to a broader audience. A focus on delivering research findings that are reproducible, or even re-usable has been proposed as one way of achieving such accessibility goals. In this work, we present the LiveDocs initiative, an effort of the "Collaborative Research Center 1456 - Mathematics of Experiment" on tackling common issues of reproducibility and re-usability in scientific publications. The LiveDocs initiative is proposed as a concept alongside a collection of methods that enable scientists to provide research findings under an interactive development environment. This environment allows users from a broader audience to easily reproduce research findings by re-running scripts, for instance, those that generate figures, tables, and other elements from scientific publications. Moreover, LiveDocs also allow the audience to interact with code and data in such environments, thus allowing users to explore algorithms, datasets and software interfaces. This directly lowers the barriers to access and comprehend research methods and findings, which facilitates more scientific exchange and fosters knowledge advancement.*

## Introduction

Open Source development, Open Access Research, and Open Research Data - often summarized as Open Science - are recurring ideals that have yet to be fully integrated into the everyday practices of scientific research.

The reuse of scientific software and datasets offers advantages for the scientific community: Researchers can build on existing solutions, yielding shorter development times; scientists offering the source code and data benefit when their software, datasets and experiments can be verified and further developed by other researchers; and society as a whole benefit when innovations are created from existing software and data. However, exploiting such benefits and enabling reuse can only succeed if scientists actively disclose their code and data.

Despite the technical availability of code and data being important for reproducible research, only making it available can be unsatisfactory. In many cases, comprehensible descriptions of execution environments, execution instructions and further explanations are crucial to make use of technically available resources. This additional information can consist of simple scripts and definition files (that may even be machine-readable) and ideally allows for perfect repeatability of computational results. An added value may be given if descriptions and explanations go beyond a level of merely technical reproducibility, but also explain the conception of methods and data, display ways to work with the provided resources, explain the research context or results or give additional, possibly simplified examples.

In response to these challenges, we propose a concept to facilitate the accessibility, reproducibility and re-usability of scientific data and source code, in the form of *LiveDoc*s (Live Documents). A LiveDoc is envisioned as a remotely executable encapsulation of development environments containing all the source code and datasets needed to reproduce scientific findings with ease. Within these LiveDocs users can easily engage with research findings, interact with scripts, and explore data sets. Such a medium is a valuable asset when used for the just-in-time reproduction of research results for any interested party. Moreover, further explanations, illustrations and examples can be delivered alongside the research findings improving the comprehensibility of research methods and findings for a broader audience.

Within the subproject "Infrastructure for Exchange of Research Data and Software" (INF) of the DFG-funded "Collaborative Research Center 1456 - Mathematics of Experiment"[1], the LiveDocs initiative has been developed. With this manuscript, we describe the conceptual foundations of LiveDocs, the technologies behind them and present two showcases. Finally, we close with a discussion on reproducibility, re-usability and accessibility and the contribution LiveDocs can make in that regard.

---

[*]Corresponding author: p.klein@math.uni-goettingen.de

[1] `https://www.uni-goettingen.de/en/628179.html`

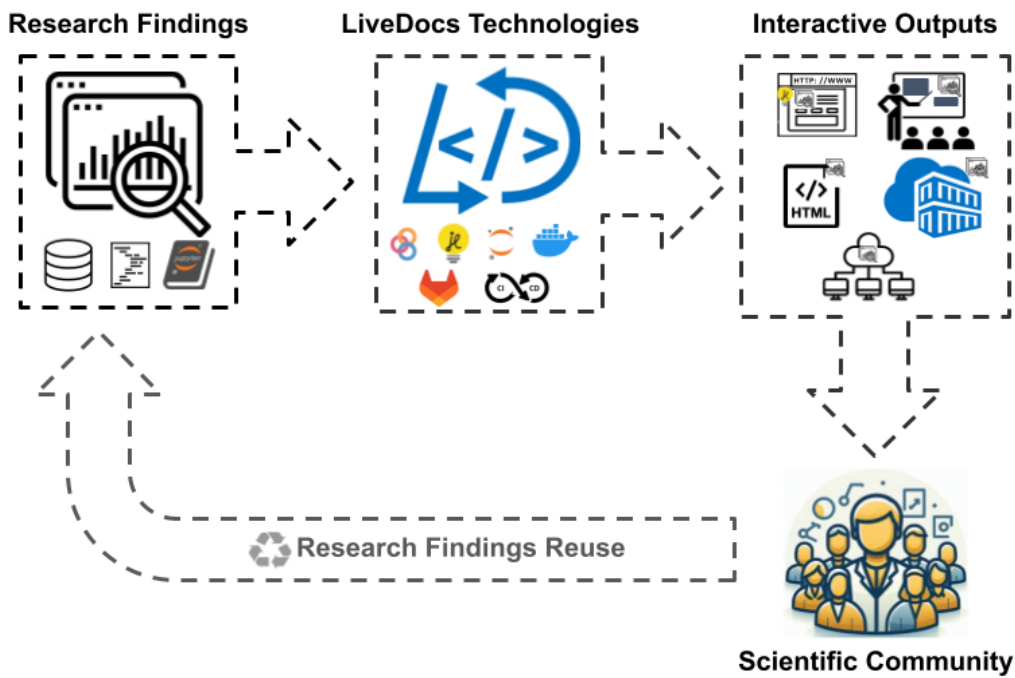

**Figure 1:** *The pipeline of a LiveDoc. A collection of research findings, in the form of a repository allied with technologies such as* `Binderhub`*,* `JupyterLite` *and* `Docker`*, allows scientists to create interactive outputs from their research findings and methods, which allows the scientific community to interact with such findings and methods, therefore facilitating the understanding of the research.*

## The LiveDocs Concept

The specification of reproduction instructions and the creation of additional explanatory content for journal publications or similar outlets for a broader audience incurs a significant investment of time and resources for scientists. This is why it is often refrained from besides its potential benefits. To lower the barrier and streamline the creation of LiveDocs for researchers, we provide means for the technical deployment with a *LiveDoc Template*. The LiveDoc Template is designed to facilitate the conversion of a collection of research data (e.g. a repository containing scripts, data sets, figures, etc.) into an accessible and interactive environment, a *LiveDoc*, so that the researcher can focus on scientific content instead of spending resources on the technical challenges.

A recommended starting point for a LiveDoc creation involves utilizing a collection of data representing the findings of a scientific publication. This collection contains scripts, data and any supplementary material needed for replicating the findings. It is then stored in a repository and made available to the general audience (Figure 1, top left). Besides making this collection available, which by itself does not ensure the reproducibility of findings, other factors underlying the data and scripts may impact its reproduction, such as specific configurations for the development environments.

These configurations may include tasks such as installing libraries, setting system environment variables, and ensuring compatibility among installed library versions. While setting up an environment for running shared code might be intuitive for those closely involved in the publication, this intuition is often temporary and diminishes over time and scientific distance. External users, and even those initially familiar with the shared code, may overlook crucial details, leading to unexpected behaviours – ranging from script failures to inaccurate results.

To circumvent such issues, the implementation of a LiveDoc through the LiveDocs template allows scientists to employ different technologies (Figure 1, top center) to encapsulate their research collections, along with all underlying requirements, into a remote development environment (Figure 1, top right), which can then be accessed by the broader audience. This reduces the amount of resources on the user's side for the access (Figure 1, bottom right) as well as on the scientist's side for making their results (more) accessible.

Note that the interactive development environments are not limited to simply reproducing findings. The idea behind LiveDocs is also to allow interaction with available code and data. This allows users, for instance, to test different scripts with the available data, or test their data with the available scripts. Such interactions promote also the reusability of the research data/scripts which, hypothetically, could incur into new research findings.

## LiveDoc Types

By incorporating LiveDocs, research projects can benefit in many ways. A researcher can, for instance, extend the explanations on topics and results from a publication beyond community-specific publications. Moreover, LiveDocs allow the addition of illustrations or ludic examples to the research findings, thus increasing the accessibility of its results and methods. Finally, sufficiently detailed LiveDocs, can be useful for teaching purposes, providing the audience with explanations and examples on the topics permeating such publications, as well as allowing interaction with such findings and methods.

With this in mind, we distinguish two types of LiveDocs: **Reproduction** and **Educational LiveDocs**.

- **Reproduction LiveDocs** are focused on allowing the users to replicate findings from an underlying scientific publications, and to interact with such findings. Such LiveDocs are directly focused on reproducibility and reusability, assuming that the target audience has prior knowledge on the topics and can comprehend the presented results relying only on the information provided in the paper.
- **Educational LiveDocs**, on the other hand, target not only at the interaction with and reproduction of research data, methods and findings but rather placing such findings in a wider scope of knowledge and covering such knowledge within the LiveDocs. In general, such LiveDocs have additional content, beyond the content of an underlying journal publication, in the form of explanations, illustrations and examples.

Note that over the course of a LiveDoc's "lifespan", a project that initially serves as a Reproduction LiveDoc, can be extended and transformed into an Educational LiveDoc. We also hope that both Educational LiveDocs and Reproduction LiveDocs will serve as starting points for new research that could then evolve directly into a new Reproduction LiveDocs.

## Technologies for LiveDocs

LiveDocs are built upon a series of technologies which allow users to create interactive development environments from code repositories. Such technologies allow the delivery of LiveDocs in many formats ("LiveDocs Flavours"). In this section, we will approach the technologies behind a LiveDoc, and briefly summarize some key technologies that are involved in the current set of existing LiveDoc Flavours.

## Backbone Technologies

The foundation of a LiveDoc is a repository containing a collection of everything needed for the replication and reproduction of research findings (a research data collection), often a `git`[2] repository. The main platforms are `gitlab`[3] and `github`[4], but other similar platforms such as `Zenodo`[5] can also be used. In the context of the CRC 1456, we use the `gitlab` instance hosted and provided by the *Gesellschaft für Wissenschaftliche Datenverarbeitung mbH Göttingen*[6] (GWDG).

Besides the research data collection, a LiveDoc repository contains all the preparations for delivering the LiveDocs in its many available formats. Such preparations usually take the form of auxiliary files such as `Dockerfiles` and lists of library requirements and of *Continous Integration and Continuous Deployment* (CI/CD) scripts, containing all the necessary steps for setting up all desired LiveDocs Flavours.

With our current set of LiveDocs, there is a focus on the three programming languages which are natively compatible with the `Jupyter Project`[7], namely `julia`[8], `Python`[9] and `R`[10]. Although `Jupyter` offers support for other languages and alternatives to `Jupyter` exist, as of now LiveDocs have been realized only for those languages, since they are the most used open-source languages within the scope of the subprojects composing the CRC 1456, which is the project from which the LiveDocs initiative originated.

Despite the setup described above being sufficient to replicate findings locally, considering, for instance, a well-defined repository in one of the predefined languages supported by `Jupyter` and a local instance of `Jupyter Server`, it does not suffice when the aim is to make the replications of findings (conveniently) accessible. Configuring a `Jupyter` server and installing all dependencies in the correct version (even if they are listed in a configuration file) might be a non-trivial and always time-consuming task for a wider audience, which might discourage them from considering using the respective LiveDoc. Therefore, we introduced alternative ways to *conveniently* run LiveDocs. These different *LiveDocs Flavours* are discussed next.

---

[2] https://git-scm.com/
[3] https://gitlab.com/
[4] https://github.com/
[5] https://zenodo.org/
[6] https://gitlab.gwdg.de/
[7] https://jupyter.org/
[8] https://julialang.org/
[9] https://www.python.org/
[10] https://www.r-project.org/

| Deployment | interactive | dependencies | ressources | private access | persistent changes | scalability w.r.t. users |
|---|---|---|---|---|---|---|
| local installation | ✅ | 📝 | 🖥 | 🔒 | ✅ | ⬆ |
| static HTML | ❌ | — | — | 🔒 | ❌ | ⬆ |
| container 🐳 docker | ✅ | 🐳 | 🖥 | 🔒 | ✅ | ⬆ |
| MyBinder 🔘 binder | ✅ | — | ▼ | — | ❌ | OK |
| self-hosted BinderHub | ✅ | — | OK | 🔒 | ❌ | ▼ |
| JupyterLite | ✅ | — | 🖥 | 🔒 | ❌ | ⬆ |

✅ yes  ❌ no  — N/A  🔒 restricted access possible  ⬆ very good  📝 local requirements  🖥 local resources  🐳 Docker  ▼ low  OK limited

**Figure 2:** *The contents of a LiveDoc Project repository can be deployed in different flavours. In this table, we give a rough overview of the advantages and disadvantages. However, note that the assessment in the table only gives a simplified view.*

## LiveDocs Flavours

LiveDocs Flavours are different deployment formats for LiveDocs. In each format, we aim at delivering the LiveDocs' contents ensuring that the conent is accessible to the user with the a small amount of intervention from the user's side. Moreover, we attempted to provide the broader audience with a wide range of flavours in an attempt to cover as many use cases as possible, with each LiveDoc Flavour having its own advantages. In this scenario, a single LiveDoc can have many flavours covering different needs of the audience, cf. Figure 2 for a rough overview of possibilities.

In the following subsections, we will briefly mention and summarize the properties of some possible LiveDoc Flavours, skipping the trivial flavour of a local installation.

**Binderhub:** `Binderhub`[11] is a cloud service based on `Kubernetes`[12] and `repo2docker`[13] which allows the creation of interactive (virtual) development environments based on code repositories. `Binderhub` creates a `JupyterHub` instance for a LiveDoc repository and can utilize various `Jupyter` extensions, including `Voilà`[14]. `Binderhub` LiveDocs instances need to be run in a backend server in which a `Binderhub` server is set up. Users can either rely on open-access solutions, such as `MyBinder`[15], or self-hosted solutions, such as the public CRC 14546 Binderhub Server[16].

With `Binderhub` LiveDoc instances, scientists can present research findings in the form of a remote (virtual) development environment in which users can interact with code and data by altering and re-running portions of code, and by uploading their own script and data to such environments. Due to their transient nature, `Binderhub` development environments are automatically wiped clean after use and no alterations are made to the original repositories.

**Containerization:** `Docker`[17] serves as a comprehensive containerization platform for the development, shipping, and execution of applications, allowing users to decouple applications from infrastructure. This separation facilitates the deployment of applications in diverse environments. In the context of LiveDocs, `Docker` emerges as a viable alternative for converting LiveDoc repositories into self-contained `Docker` images. These images can operate within independent containers, fostering a portable and interoperable environment.

With `Docker` LiveDoc instances, scientists are allowed to present broad audience with a standalone image of a LiveDoc repository, with which users can interact by running such images in containers. Differently from other approaches, `Docker` LiveDoc instances rely on user's resources – thus also allows for execution on computing clusters – and can offer some persistence on alterations. This means that the portable development environments can be retained after use.

**JupyterLite:** `JupyterLite`[18] is a lightweight, client-side implementation of the `Jupyter` interactive computing environment using WebAssembly (`wasm`). It is designed to run entirely in the browser without the need for a server. `JupyterLite` delivers the users the familiar

---

[11] https://binderhub.readthedocs.io/
[12] https://kubernetes.io/
[13] https://repo2docker.readthedocs.io/
[14] https://voila.readthedocs.io/
[15] https://mybinder.org/
[16] http://c109-005.cloud.gwdg.de:30901/
[17] https://www.docker.com/
[18] https://jupyterlite.readthedocs.io/

`Jupyter notebook` interface for creating and sharing documents containing live code, equations, visualizations, and narrative text.

In the context of LiveDocs, `JupyterLite` is generally used in cases in which scientists want to provide the audience with an alternative to remote cloud instances such as `Binderhub`, but with a similarly effortless use setup. Due to its `wasm`-based nature, `Jupyterlite` instances can be run directly on users' computers (or even mobile devices) – within the user's browser to be precise – therefore discarding the need for external resources such as `MyBinder` or self-hosted services. This is specifically attractive in view of the scalability for many users accessing the same LiveDoc. While resources of the `Binderhub` servers would need to be shared among parallel users, `Jupyterlite` LiveDocs essentially rely only on the user's resources.

Despite its flexibility in usage, `Jupyterlite` suffers from limitations at runtime (32 bit and browser resources) and – more severe – in the setup, from the LiveDocs perspective: the desired LiveDocs need to run completely in `wasm`, i.e. all depending code needs to be compiled in `wasm`. With `Pyodide`[19], the `Python` kernel for `WebAssembly` pure `Python` packages work essentially out of the box. However, for packages or dependencies that are not available as `wasm`-compiled libraries additional effort would need to be put in order to provide them for a `JupyterLite` instance.

**Static LiveDocs:** In the context of LiveDocs, another way of representing research findings is in the form of static LiveDocs. Static LiveDocs are a simplification of the interactive LiveDocs, aiming at the easy distribution and quick assessment of research findings. The static LiveDocs are usually presented in the form of static `HTML` files, generated from `Jupyter` notebooks using `NBConvert`[20].

The main advantage of this approach is that users have a point to quickly access scripts and inspect code and figures generated with such scripts, with (almost) no local or remote resources required. The downside of this approach is that users have no interaction with code, being presented only with a "snapshot" of the Jupyter notebooks. In the context of LiveDocs, such static approaches are suited for quick assessment as a back-up deployment and as an initial contact with research findings, leading then to the usage of more sophisticated LiveDocs flavours.

## LiveDocs Templates

We have established a template repository under

`https://gitlab.gwdg.de/crc1456/livedocs/livedocs_template`

that streamlines the creation of LiveDocs by converting Jupyter-like documents into various formats, including static HTML, `Docker` containers, or `JupyterLite` instances. This initiative provides flexibility in document dissemination while ensuring consistency in the overall presentation. Moreover, in this template, scientists can also find a walk-through on the basic LiveDocs functionalities.

As for its usage, this template is intended to either act as the founding stone for a new repository, in which users will add their own research data collection, or be merged with an existing repository already containing a research data collection. In both cases, some further configuration is needed from the user's side, such as filling in the requirements list with necessary libraries and their versions (when needed).

The adoption of the template is not mandatory, and realizations of a LiveDoc can differ from the provided template.

## LiveDocs Showcases

This LiveDocs initiative with the CRC 1456 also involves the development of a centralized landing page that provides a curated and intuitive entry point into the diverse array of LiveDocs within the scope of the CRC 1456 which you can find here

`https://crc1456.pages.gwdg.de/livedocsrepo/`

The landing page outlines the showcases and highlights the breadth and depth of research covered across the CRC.

The INF project is dedicated to establishing the necessary technical infrastructure for LiveDocs and actively supports CRC 1456 research groups in their creation. This collaborative effort ensures that the CRC's research outcomes are not only assessed w.r.t. their technical reproducibility but also presented in an accessible and engaging manner, promoting knowledge exchange within and beyond the CRC. In the following, we mention two examples for LiveDocs stemming from research within the CRC, one for an Educational LiveDoc and one for a Reproduction LiveDoc.

---

[19] https://pyodide.org/
[20] https://nbconvert.readthedocs.io/

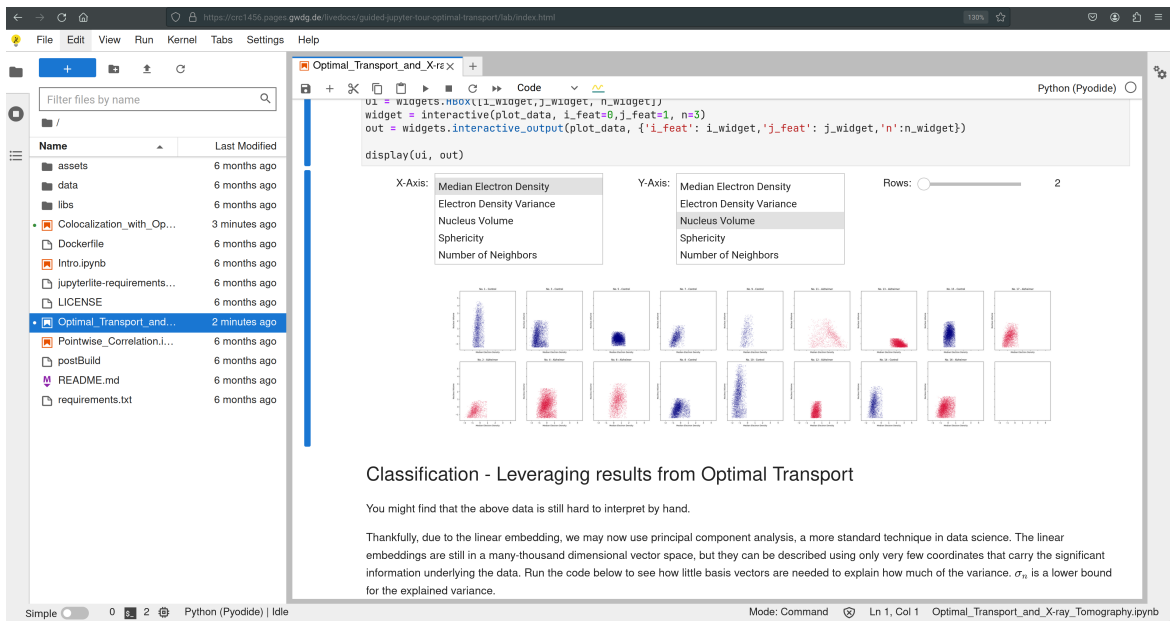

**Figure 3:** *Screenshot of a* `JupyterLite` *instance of LiveDocs "Guided Tour to Optimal Transport".*

## Guided Jupyter Tour: Optimal Transport (Educational LiveDoc)

In the first LiveDoc [1], cf. Figure 3 for a screenshot, users are presented with an introduction to the topic of Optimal Transport problems and computational approaches, with a collection of interactive notebooks that range from introductory, to "real world" examples.

The target audience for this LiveDoc is intermediate Bachelor's students in mathematics or Computer Science. Some notebooks in this collection address recent research [2] and may be interesting also for students of Master's and Doctoral programs. For audiences without mathematical background, the introductory notebooks, with interactive widgets and graphs, are recommended, since these were purposefully made accessible without much in the way of mathematical background knowledge.

The notebooks in this LiveDoc are written in `Python3` with `Jupyter` and `ipywidgets`[21] as the interactive elements. They utilize `numpy`[22] and `scipy`[23]. An open-source library for solving optimal transport problems[24] developed by Prof. Bernhard Schmitzer and collaboratos was also used. This LiveDoc was developed from May to July 2022. It builds on previous work in the form of Bachelor's theses by Fabian Fieberg and Thilo Stier.

As for the authors, the programming was made by Thilo Stier, Fabian Fieberg and Lennart Finke. The texts in the notebooks were written by Lennart Finke. This work was supervised by Prof. Dr. Cristoph Lehrenfeld and Prof. Dr. Bernhard Schmitzer. The notebooks are licensed under MIT license.

## 3d Virtual Histology Reveals Pathological Alterations of Cerebellar Granule Cells in Multiple Sclerosis (Reproduction LiveDoc)

In the second LiveDoc discussed here, LiveDoc [3], cf. Figure 4 for screenshot, users are presented with a virtual environment with all the requirements set up for reproducing the findings contained in the publication [4]. This publication investigates the structural properties of cerebellar neurons in the human granular layer, concerning multiple sclerosis utilizing X-ray tomography on post-mortem tissue.

The LiveDoc related to this publication contains three `Jupyter` notebooks with scripts related to the reproduction of findings from this publication, along with some explanatory text. Besides that, this LiveDoc repository contains the data needed for running the scripts and a collection of Optimal Transport Algorithms (Barycenter computation, OT Solver) created by Bernhard Schmitzer and the Optimal Transport Group in Göttingen[25]

## Discussion

Reproducibility is an essential goal in the current scientific method, ensuring the robustness and reliability of findings. However, concerns about the lack of reproducibility in various scientific fields have been raised in recent years [5]. In this context, research that does not provide adequate details about its methods or make data available for scrutiny hinders further efforts to replicate studies and evaluate their findings.

---

[21] https://ipywidgets.readthedocs.io/
[22] https://numpy.org/
[23] https://scipy.org/
[24] https://github.com/bernhard-schmitzer/UnbalancedLOT
[25] https://github.com/bernhard-schmitzer/UnbalancedLOT

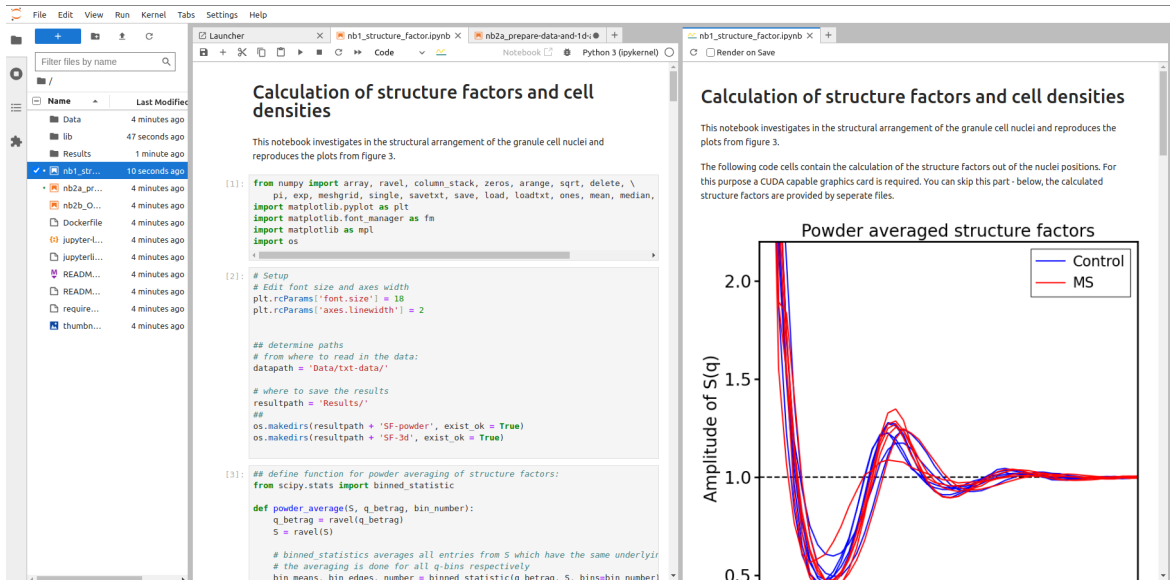

**Figure 4:** *Screenshot of a `JupyterLab` instance of LiveDocs "3d Virtual Histology Reveals Pathological Alterations of Cerebellar Granule Cells in Multiple Sclerosis", with a `Voilà`-generated condensed version consisting only of explanation, plots and widgets (without the code) on the right.*

Re-usability in academic research promotes efficiency, collaboration, and the advancement of knowledge. In this context, re-usability comes with a growing emphasis on open science practices, including sharing datasets, code, and detailed methodological information. Initiatives such as open-access publishing, data repositories, and collaborative platforms contribute to creating a more accessible and interconnected research environment. Such practices enhance the impact of research, encourage collaboration, and potentially accelerate the pace of scientific discoveries.

There is a noticeable effort in the scientific community to improve the infrastructure supporting data reproducibility and reuse. Initiatives such as the establishment of the "FAIR" principles [6] point the scientific community towards good practices for research data management. The FAIR principles state that data should be Findable, Accessible, Reusable and Interoperable, thus enhancing research data dissemination.

In the present work, we introduced LiveDocs, an attempt for research dissemination that tackles issues of reproducibility and re-usability of research findings. Such an approach is posed as one potential solution to replication problems, allowing users to conveniently assess results and methods presented in publications. Moreover, LiveDocs also offers users a pre-configured development environment in which they can, on top of reproducing findings, interact with code and experiment with new approaches over the presented data, or try new data using the available scripts. Such interactions offer the potential of moving research topics to a further point.

## Conclusion

We presented the idea behind the LiveDocs initiative and presented the currently applied technologies to render them conveniently realizable for many cases. LiveDocs contribute to the advancement of Open Science by enhancing the transparency and accessibility of scientific findings, fostering collaboration and knowledge exchange.

## Acknowledgements

The authors acknowledge funding by DFG SFB 1456 project 432680300. Furthermore, we thank the authors of the LiveDocs mentioned in this article for their participation in the LiveDocs initiative.